\documentclass[twocolumn]{aastex631}
\shorttitle{Ly-$\alpha$ field Reconstruction}
\shortauthors{Jacobus, Harrington, \& Luki\'c}
\graphicspath{{./}{figures/}}
\usepackage{amsmath}
\usepackage[cal=Euler]{mathalfa}
\usepackage{listings}
\usepackage{bigints}
\usepackage{lipsum}
\setlipsum{%
  par-before = \begingroup\color{gray},
  par-after = \endgroup
}

\usepackage{booktabs}

\begin{document}

\title{Reconstructing Lyman-$\alpha$ Fields from Low-Resolution Hydrodynamical Simulations \protect\\ with Deep Learning}

\author{Cooper Jacobus}
\affiliation{Lawrence Berkeley National Laboratory, Berkeley, CA 94720, USA}
\affiliation{Department of Astronomy, University of California, Berkeley, CA 94720, USA}

\author{Peter Harrington}
\affiliation{Lawrence Berkeley National Laboratory, Berkeley, CA 94720, USA}

\author{Zarija Luki\'c}
\affiliation{Lawrence Berkeley National Laboratory, Berkeley, CA 94720, USA}



\begin{abstract}

%To best understand the distribution of matter in the universe and study the formation of the cosmos, researchers hope to compare statistical observations of galaxy distributions with simulated universes. 

Hydrodynamical cosmological simulations are a powerful tool for accurately predicting the properties of the intergalactic medium (IGM) and for producing mock skies that can be compared against observational data. However, the need to resolve density fluctuation in the IGM puts a stringent requirement on the resolution of such simulations which in turn limits the volumes which can be modelled, even on most powerful supercomputers.  In this work, we present a novel modeling method which combines physics-driven simulations with data-driven generative neural networks to produce outputs that are qualitatively and statistically close to the outputs of hydrodynamical simulations employing 8 times higher resolution.  We show that the Ly-$\alpha$ flux field, as well as the underlying hydrodynamic fields, have greatly improved statistical fidelity over a low-resolution simulation.  Importantly, the design of our neural network allows for sampling multiple realizations from a given input, enabling us to quantify the model uncertainty. Using test data, we demonstrate that this model uncertainty correlates well with the true error of the Ly-$\alpha$ flux prediction. Ultimately, our approach allows for training on small simulation volumes and applying it to much larger ones, opening the door to producing accurate Ly-$\alpha$ mock skies in volumes of Hubble size, as will be probed with DESI and future spectroscopic sky surveys.

\end{abstract}


%% cross-references to sections, equations, tables, and Figures.

\section{Introduction} \label{sec:intro}

The neutral hydrogen in the intergalactic medium imprints a characteristic pattern in the absorption spectra of quasars, known as the ``Ly-$\alpha$ Forest''.  The Ly-$\alpha$ forest is a powerful probe of the thermal state of the intergalactic medium, and also of cosmology. The distribution of the Ly-$\alpha$ forest lines depends on the distribution and properties of the intervening gas, which is affected by the background of UV photons and the history of cosmological structure formation (for a recent review, see~\cite{McQuinn2016}). 

Two main ways to infer cosmological parameters from the Ly-$\alpha$ data are 1D power spectrum and Baryon Acoustic Oscillations (BAO). In the case of the 1D power spectrum (P1D), relatively small-scale correlations (down to $1 h^{-1}$Mpc) along the line of sight are used to constrain cosmology~\citep{Croft1999, McDonald2005, PDB2013, Chabanier2019a, Naim2022}. This probe was especially proven effective for constraining the sum of neutrino masses~\citep{Seljak2005, PDB2015a, PDB2015b, PDB2019}, warm dark matter models~\citep{Viel2005, Viel2013, Baur2017, Yeche2017}, fuzzy dark matter~\citep{Irsic2017, Armengaud2017} or interacting dark matter models~\citep{Dvorkin2014, Xu2018, Garny2018}.  BAO on the other hand, is a large-scale measurement probing matter clustering on scales of $\sim 100 h^{-1}$ Mpc, and providing geometrical constraints on the expansion rate and the angular diameter distance~\citep{Slosar2013, Bautista2017, deSainteAgathe2019, DuMas2020}.

\begin{figure*}[t]
  \centering
  \includegraphics[width=\textwidth]{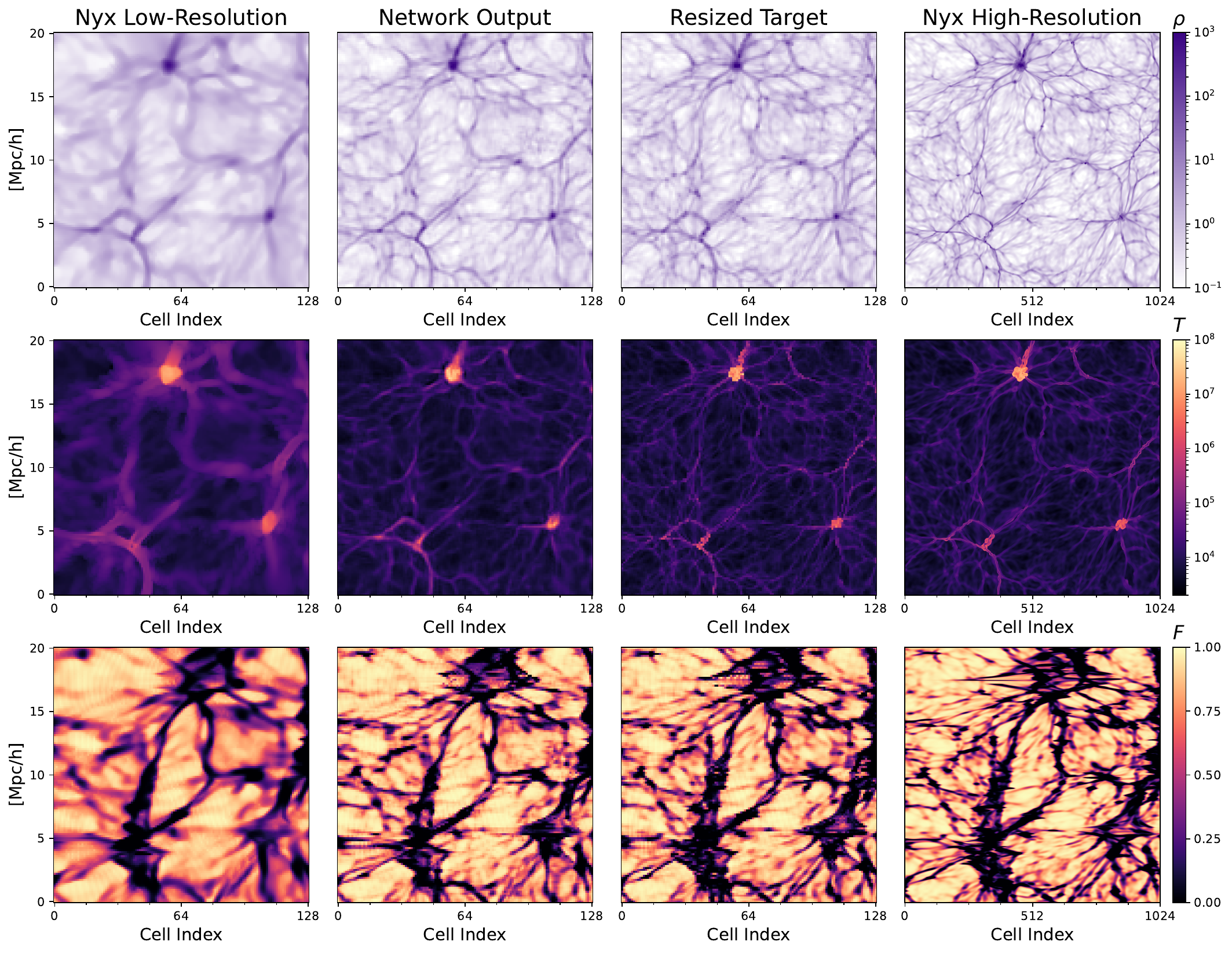}
  \caption{Comparison of simulated and predicted Hydrodynamic fields and Lyman-$\alpha$ flux. The first column is the low-resolution simulation fed as input to our model, which is trained to reconstruct features from the high-resolution simulation (fourth column) using a version of it resized to an 8x coarser grid as the training target (third column). Our model predictions are shown in the second column, demonstrating clear improvement over the low-resolution simulation.} 
  \label{fig:all_comp}
\end{figure*}

However, the real potential of the Ly-$\alpha$ forest lies in using the full shape of the three-dimensional power spectra~\citep{McDonald2003}. This probe combines cosmological information from both large and small scales, radically expands the number of available modes, and breaks some of parameter degeneracies due to capturing correlations both along and perpendicular to the line of sight (see, for example~\cite{Rorai2017}). This in turn offers the possibility of significantly improving constraining power on cosmological parameters compared to BAO or P1D. The measurement is difficult and has not been performed so far, because it requires both surveying large volume and having large number of lines of sight.  However, the ongoing Dark Energy Spectroscopic Instrument~\citep{DESI2016} will be the first survey to enable measurement of P3D~\citep{FontRibera2018} due to its quasar density of $\sim$60 QSO/deg$^2$~\citep{Chaussidon2022}, or approximately 3 times denser than previous surveys (BOSS, eBOSS). In the future, the Stage V Spectroscopic Facility, which is currently being scoped by the cosmological community~\citep{snowmass2022}, will further improve the precision of such measurement.

Inferring parameters from P3D requires not only precise observations, but also accurate theoretical modeling, which is difficult because there is no analytic solution for the small-scale evolution of the (baryon) density fluctuations. In order to precisely model the state of the IGM, it is instead necessary to approach the problem numerically. In this respect, hydrodynamic cosmological simulations have led to a consistent description of the IGM in the framework of large-scale structure formation~\citep{Cen1994}. However, such simulations are computationally expensive, as they must retain the high spatial resolution of $\sim 20 h^{-1}$kpc, which is necessary to resolve density fluctuations in the IGM~\citep{Lukic2014, Walther2021, Chabanier2023, Doughty2023} to accurately capture the Ly-$\alpha$ signal, despite the fact that observatories like DESI are measuring fluctuations in Ly-$\alpha$ flux only at scales $\gtrsim $1 Mpc $h^{-1}$ \citep{Karacayli2023, Ravoux2023}.  Furthermore, constraints on memory limits the size of volumes that can be run in high-resolution simulations, as for example a 1$h^{-1}$Gpc simulation would require $\sim50,000^3$ computational elements, far beyond computational resources in the foreseeable future (note that this requirement applies to both Eulerian and Lagrangian codes, see~\cite{Chabanier2023}).

In addition to P3D, large volumes enable cross-correlating Ly-$\alpha$ and other tracers of matter fluctuations.  Since the forest power spectrum is sensitive to the small-scale matter fluctuation whereas CMB lensing convergence measures the total matter fluctuation along the line of sight, combining the two~\citep{Vallinotto2009, Vallinotto2011, Chiang2018, Dash2021} allows us to probe the nonlinear structure formation at large scales in a redshift range still poorly tested.  Correlating galaxy positions and their Ly-$\alpha$ environments are another promising research avenue~\citep{Font-Ribera2013, Bielby2017, Mukae2017, Momose2021a, Momose2021b} for tracing matter fluctuations, as galaxy formation occurs in preferentially more dense environments, while the forest mostly comes from less dense regions.  This synergy between the two can also significantly improve the quality of density and initial conditions reconstruction~\citep{Horowitz2021}.

Recently, machine learning, and more specifically deep learning, has shown remarkable potential to assist in solving highly non-linear problems like this by serving as a surrogate model for complex physics simulations (for a recent review of many deep learning applications in cosmology, see~\cite{Huertas-Company2023}).
Methods based on deep learning have been used to replace full simulations~\citep{He2019, Mustafa2019, Feder2020, Giusarma2023}, to substitute expensive hydrodynamical solvers by inferring gas properties from N-body simulations~\citep{Troster2019, Thiele2020, Dai2021, Bernardini2022, Harrington_2022, Horowitz_2022, boonkongkird2023lyalnet}, or going even deeper into numerical algorithms, to replace chemical network solvers inside hydro simulations~\citep{Fan2022}. 

Relatedly, a number of works have employed deep learning to super-resolve the fields from coarse simulations, synthesizing fine-scale details without incurring large computational costs. This has been demonstrated successfully in the realm of N-body simulations and modeling of dark matter halos and small-scale galaxy formation physics \citep{li2021ai, ni2021ai, schaurecker2021super}. While these results are promising, directly applying super-resolution approaches in hydrodynamic simulations for Ly-$\alpha$ poses a number of practical challenges. For one, as mentioned previously, the simultaneous requirement of high resolution ($\sim 20$ kpc) and large volumes (1$h^{-1}$ Gpc) needed for accuracy in Ly-$\alpha$ means that direct super-resolution models would generate overwhelmingly large outputs. This drawback is compounded by the fact that in order to converge on the Ly-$\alpha$ statistics at 
$\sim 1h^{-1}$ Mpc scales (roughly the observational limit), the simulations must be run at $\sim 20h^{-1}$ kpc resolution, even though the hydrodynamic outputs themselves are not necessarily needed on scales finer than $\sim 1h^{-1}$ Mpc for analysis. It is therefore of interest to identify deep learning approaches which might act similarly to super-resolution, providing reconstructions of the physics from fully-resolved simulations, while side-stepping these practical constraints to enable deployment in survey-scale cosmological volumes.

In this work, we aim to advance the state-of-the art in deep-learning-assisted modelling of the Ly-$\alpha$ forest and demonstrate a methodology for producing a mock sky at high-resolution over volumes comparable to that of DESI. We develop a novel approach combining conditional generative adversarial networks \citep{Goodfellow2014,mirza2014conditional} with paired sets of hydrodynamic simulations, which allows us to take cheaper, coarse-resolution simulations and correct their outputs to more closely resemble the result from high-resolution, fully-converged simulations that would be infeasible to run on very large volumes. We carefully design our network and training task to ensure our approach is lightweight in terms of compute and memory requirements, while still providing high-fidelity results that greatly improve on the low-resolution simulations in terms of qualitative and statistical fidelity. Furthermore, our model can be sampled repeatedly to produce multiple realizations given a single input, and we find that these provide Monte Carlo estimates of model uncertainty which correlate well with prediction error. Finally, we also explore the application of our approach to modeling not just the Ly-$\alpha$ forest, but also the underlying hydrodynamic fields of interest such as baryon density and temperature, demonstrating yet again significant improvements over the low-resolution simulations.

This paper is organized as follows.  We first discuss the simulations we use to train our network and the methods used to compute Ly-$\alpha$ forest from them in the Section \ref{sec:simulations}. Then we describe the design and novelty of our model architecture and training incentives in Section \ref{sec:model}. Our results are presented in Section \ref{sec:results}, where we first discuss the accuracy of our model on the Hydrodynamic field reconstructions and then the statistical accuracy of the observable Ly-$\alpha$ absorption features derived from them. Finally, we present conclusions in Section \ref{sec:conclusion}.

\section{Simulations} \label{sec:simulations}

We assemble our training and testing data from pairs of simulations run by the publicly available \texttt{Nyx}\footnote{\url{https://amrex-astro.github.io/Nyx}} code \citep{Almgren_2013, Sexton2021}. The code follows the evolution of dark matter using collisionless self-gravitating Lagrangian particles which are coupled to baryonic ideal gas modeled on a uniform Cartesian grid. Although \texttt{Nyx} is capable of employing adaptive mesh refinement (AMR), we do not make use of this functionality here as the Ly-$\alpha$ signal spans the vast low-density volumes and is fully absorbed close to halos where AMR would be relevant.

To model the Lyman $\alpha$ forest, \texttt{Nyx} models primordial gas composition of Hydrogen and Helium, and follows the abundance of six species: neutral and ionized hydrogen, neutral, once, and twice ionized helium, and free electrons. For these species, all relevant atomic processes -- ionization, recombination, and free-free transitions are modeled. Heating and cooling source terms are calculated using a sub-cycled approach in order to avoid running the whole code on a short, cooling timescale.  It is assumed that all gas elements are optically thin to ionizing photons, and that the ionization state can be fully described by a uniform and time-varying UV background radiation field~\citep{Onorbe2017}. This type of simulation is very common in studies of the intergalactic medium and is used as a forward model in virtually any recent inference work using the Lyman alpha power spectrum\citep{Boera2019, Walther2019, PDB2020, Rogers2020, Walther_2021}.  Finally, the agreement between the Nyx code which employs the Eulerian approach to hydrodynamics and the SPH method was recently demonstrated in \cite{Chabanier2023}.

The simulations are initialized at $z = 200$, using the Zel’dovich approximation~\citep{zeldovich1970}. Transfer functions were generated with the Boltzmann solver code CLASS~\citep{class2011}. The cosmological parameters are set according to the Planck-2016~\citep{Planck2015} model: $\Omega_b = 0.0487$, $\Omega_m = 0.31$, $H_0 = 67.5$, $n_s = 0.96$ and $\sigma_8 = 0.83$.  We simulate volumes of 80 $h^{-1}$Mpc on a side, which is the same size that will be used in DESI 1D power spectrum modeling \citep{Walther_2021}.  We produce two pairs of such simulations, where each pair has a high- and low-resolution run initialized from the same random initial conditions. The high-resolution simulations have 4096$^3$ elements (particles and grid cells) and are thus fully converged and percent-level accurate, while the low-resolution simulations are much coarser with only 512$^3$ elements. All simulation runs have identical cosmology, UV background, and all numerical parameters, and the two pairs of high- and low-resolution simulations differ only in the choice of random realization.  This enables the model to be trained on one pair of simulations and tested on another.

\begin{figure*}[t]
  \centering
  \includegraphics[width=\textwidth]{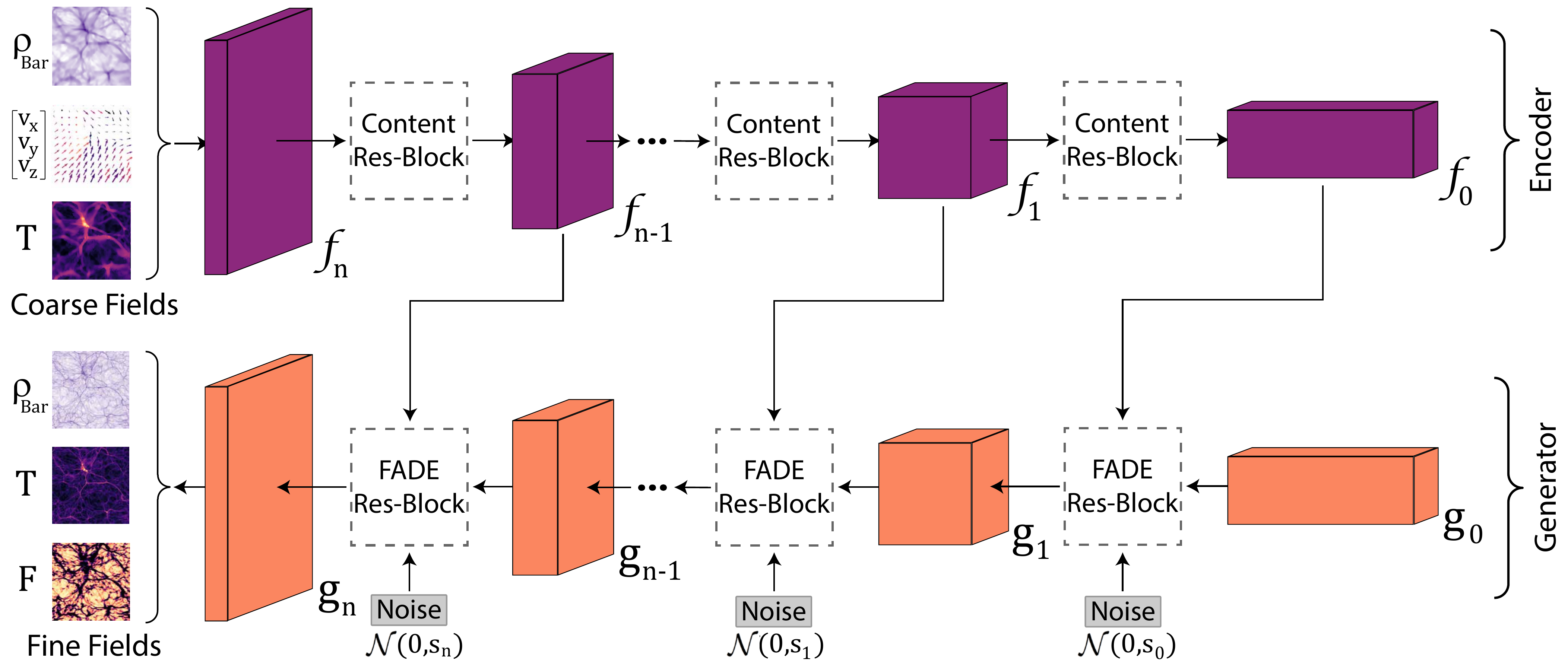}
  \caption{Schematic of our model architecture. We pass as input the low-resolution density, velocity, and temperature fields. These are processed by a convolutional encoder which extracts features at multiple scales, which are then used to by the generator to reconstruct the target density, temperature, and Ly-$\alpha$ flux fields. We inject learned multi-scale noise in the generator, enabling uncertainty quantification by making our predictions stochastic.}
  \label{fig:flowchart}
\end{figure*}

\subsection{Derived Fields} \label{subsec:derived}

We use the software suite \texttt{gimlet} \citep{Friesen2016} to derive the optical depth and flux fields from out hydrodynamic fields. The optical depth, $\tau$, for Ly-$\alpha$ photon scattering is given by: 

% optical depth -> flux
\begin{equation}
\tau_{\nu} = \int n_{X} \sigma_{\nu} \textrm{d}r
\end{equation}

%$\nu$ is the frequency
\noindent
where $n_{X}$ is the number density of species $X$ (HI in our case),  $\sigma_{\nu}$ is the cross-section of the interaction, and d$r$ is the proper path length element. We assume a Doppler line profile which results in the following optical depth:

\begin{equation}
\tau_{\nu} = \bigl(\frac{\pi e^2}{m_e c} f_{12} \bigr) \bigintsss  \bigl(\frac{n_{X}}{\Delta \nu_D} \bigr) \biggl(\frac{\textrm{exp}[-(\frac{\nu - \nu_0}{\Delta \nu_D})^2]}{\sqrt{\pi}} \biggr) \textrm{d}r
\end{equation}

where $\nu_D = (b/c) \nu_0$ is the Doppler width with the Doppler parameter $b = b_{\textrm{thermal}} = \sqrt{2 k_B T / m_H}$, and $f_{12}$ is the upward oscillator strength of the Ly-$\alpha$ resonance transition of frequency $\nu_0$, $e$ is the elementary
charge, $m_e$ is the electron mass, and $c$ is the speed of light \citep{Lukic2014}. In the optically thin limit (neglecting scattering
effects), $\tau$ becomes an entirely local quantity.
Finally, Lyman-$\alpha$ flux is given by:

% optical depth -> flux
\begin{equation} \label{eq:flux}
F = \textrm{exp($-\tau$)}
\end{equation}

\subsection{Producing Target Fields}
\label{sec:downsampled_targets}

The goal of this work is to enable ``correcting'' the output of a low-resolution hydrodynamical simulation, without increasing its memory footprint.  In other words, we do not want to super-resolve such simulation adding elements of finer resolution, but to improve the accuracy of the already resolved scales. We do this in order to enable modeling of survey-level volumes.  For instance, a volume that measures $600$ $h^{-1}$Mpc on a side --- which would be on the lower-end of survey volumes --- would require fields of size $30720^3$ to represent in high resolution, if we were to super-resolve it. Clearly, storing outputs of such sizes is not a viable approach in the foreseeable future.

Thus, we must first down-sample each field from a 4096$^3$ simulation down 512$^3$, matching the memory footprint of the low-resolution simulation.  This is not a trivial step, because we must not only make accurate coarse representations of fine features (which are represented on grids with 8x more resolution elements) but also faithfully preserve the relevant correlations of each field (especially the probability density functions and power spectra). For the target flux, we can simply subsample without introducing statistical artifacts. But for the hydrodynamic fields that span many orders of magnitude, we find that a special recursive blur and subsample method works best in resizing the high-resolution fields down by a factor of 8 while preserving statistical fidelity, which we discuss in detail in Appendix \ref{sec:downsampling}.  We use this resized volume as the training target for our model, and show sample slices from the high and low resolution simulations along with the resized training target and model outputs in Fig. \ref{fig:all_comp}.

\section{Model Design} \label{sec:model}

Modeling the hydrodynamic fields from high-resolution simulations on a coarse grid presents an interesting design challenge for a deep learning model. The neural network must model data which exhibits sharp, potentially pixel-scale features, and the mapping from coarse to fine physical fields is inherently uncertain in a non-trivial, data-dependent manner (as in the more conventional direct super-resolution task, e.g. \cite{ning2021uncertain}). To address these challenges, we adopt a generative modeling approach, heavily inspired by high-resolution image-to-image mapping research \citep{TSIT} as well as the StyleGAN \citep{karras2019stylebased, karras2020analyzing} family of models. Our model architecture can be viewed as a conditional GAN which models the distribution of (resized) high-resolution hydrodynamic fields, given the corresponding fields from the low-resolution simulation. Our design facilitates modeling both sharp, localized features as well as sampling multiple realizations of the target field from a single input, allowing direct estimation of model and data uncertainty via ensemble predictions with a single set of trained network weights.

\subsection{Multi-scale convolutional architecture} \label{subsec:TSIT}

Since our training and inference targets have already been downsampled (see Section \ref{sec:downsampled_targets}), our model architecture does not need to produce additional resolution elements, so the network outputs can be represented on a spatial grid of the same size as the input. Thus, from a machine learning perspective, our task bears a strong resemblance to the general idea of image-to-image mapping \citep{isola2018imagetoimage}, which was first attempted with architectures like the ubiquitous U-Net \citep{ronneberger2015unet} trained with reconstruction ($\mathcal{L}_1$) and adversarial loss functions. In subsequent years the computer vision community has made progress advancing the models used in this area to handle higher-resolution data via multi-scale adversarial training \citep{wang2018pix2pixHD} as well as generating more detailed and perceptually realistic outputs via improved incorporation of input features within the network \citep{park2019semantic, TSIT}. We base our model architecture on one such successful model in this area, TSIT \citep{TSIT}, which has additionally been demonstrated to show promise in scientific applications \citep{duncan2022generative}. TSIT is a fully convolutional model, which can be trained on smaller volumes cropped from a simulation and then deployed on the larger volume during inference, which loosens memory requirements for training \citep{Harrington_2022}.

We show a schematic of our model in Figure \ref{fig:flowchart}. The primary components are a pair of encoder and generator networks, which are comprised of convolution-based residual blocks \citep{ResBlocks} assembled in a symmetric manner with downsampling/upsampling between each block in the encoder/generator respectively. The encoder network takes as input the coarse hydrodynamical fields from a low-resolution simulation and extracts features from them at multiple spatial scales via successive residual blocks (ResBlocks) and downsampling operations. The extracted multiscale features are able to encode fine-level details as well as long-range information, both of which are needed to synthesize a physically realistic prediction that is consistent with the low-resolution input. Each ResBlock in the encoder consists of three convolutional layers, one of which forms a learned residual skip connection.

\begin{figure}[t]
  \centering
  \includegraphics[width=\columnwidth]{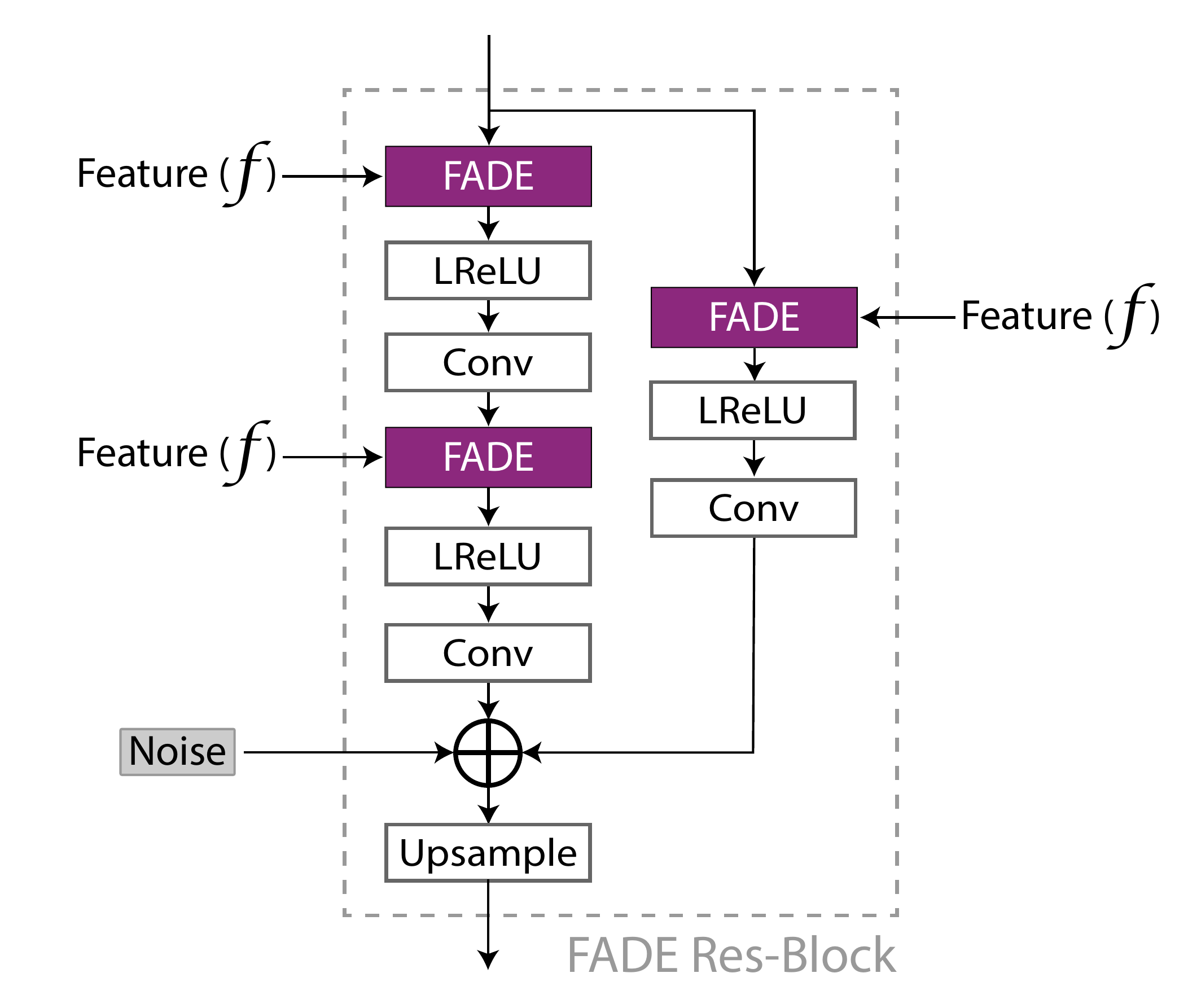}
  \caption{The FADE residual block submodule in the generator. The FADE module performs element-wise denormalization by modulating the normalized activation using a learned affine transformation. Adapted from Figure 3 in \cite{TSIT}}
  \label{fig:FADE}
\end{figure}

The generator network operates in a coarse-to-fine direction, incorporating extracted features from the encoder as well as sampled latent variables (given by Gaussian noise) at progressively larger spatial scales to form the final prediction. In the following section, we describe in detail how the generator samples latent variables via noise injection, but here we must first describe how the encoder features are incorporated into the generator. This is done using the FADE ResBlocks, proposed in the original TSIT work \citep{TSIT}, which we illustrate in Fig. \ref{fig:FADE}. The FADE ResBlocks are largely the same as the standard ResBlocks in the encoder network, except for the addition of FADE modules which take extracted features $f_i$ from the encoder and use them to perform adaptive instance normalization. This design, inspired from style transfer literature, allows the generator to flexibly incorporate multiscale features and use them to synthesize multiple outputs from a single input easily. We refer the reader to \cite{TSIT} for further discussion on the advantages of the feature-wise adaptive normalization approach.

\subsection{Latent Variable Noise Injection} \label{subsec:noise}

A major goal of our work is to develop a non-deterministic model, such that we can generate multiple predictions of the high-resolution targets given a single input. This allows us to make ensembles of predictions which in aggregate give a sense of the combined model and data uncertainty associated with our neural network approach. The TSIT work does explore multiple ways of achieving this, mainly through either sampling a noise vector at the bottom of the generator network (similar to a standard GAN), as well as via an additional encoder into a latent space which is structured like that of a variational auto-encoder (VAE).

Through initial experimentation with these methods, we found that neither method was sufficient to produce a statistically significant amount of variation in the network outputs. We suspect that part of the problem with these approaches is that they only introduce noise or variability at the ``bottom'' end of the generator, which represents features at the coarsest spatial scales. In our data, however, it is likely that much of the variation and uncertainty inherent to our coarse-to-fine mapping actually lies in the finest spatial scales. Thus only introducing noise at the coarsest scales may not give the generator network adequate means to express a full range of variability in outputs (see further discussion of this concern in Section 3.2 of \cite{karras2019stylebased}).

To address this concern and produce more stochastic behavior, we leverage multi-scale latent variables which inject additive Gaussian noise into the generator at all feature resolutions, similar to the StyleGAN family of models \citep{karras2019stylebased}. With this approach, the generator is far less dependent on only the noise at the bottom of the network to produce variation, so much so that previous works have produced equally high quality and variable outputs even when simply replacing the bottom-level noise input with a constant, learnable tensor \citep{karras2019stylebased, karras2020analyzing} or simply a downsampled version of the input features \citep{park2019semantic}. We use the latter in this work, and then rely on the multiscale noise and FADE blocks in the later stages of the generator to produce physically realistic variation in our network outputs.

As the correct scale of noise to inject cannot be known \textit{a priori}, and indeed the correct scale may change throughout training as the internal network weights and activations drift from random initialization towards their converged values, we choose to let the network learn the proper noise scale as a trainable parameter. Specifically, for each noise injection $z_i$ at a given stage in the generator $g_i$, we sample a tensor of uncorrelated Gaussian noise from $\mathcal{N}(0,s_i)$, where $s_i$ is learnable. The tensor of noise is directly added to the output of the FADE ResBlock in $g_i$, and the result is then passed to the subsequent generator stage $g_{i+1}$. At the beginning of training, we initialize $s_i$ at all stages of the generator to be sampled from $\mathcal{N}(1,0.05)$. Overall, our noise injection approach is most similar to that of \cite{duncan2022generative}, who found it to work well in producing physically realistic variability in generative weather forecasting.

\subsection{Normalization} \label{sec:training}

The hydrodynamic fields produced by \texttt{Nyx} span several orders of magnitude. Features in the baryon density and temperature fields are most easily interpretable in log-space. We follow \cite{Harrington_2022} and use the following functions to normalize the hydrodynamic physical fields before passing them to the model: 

% Hydro normalizations
\begin{equation}
\begin{gathered}
  \hat{\rho} = log(\rho)/14 \\
  \hat{v} = v / (9 \times 10^7 \textrm{cm/s}) \\
  \hat{T} = log(T)/8 - 1.5 
\end{gathered}
\end{equation}

\noindent
where $\rho$ is baryon over-density, $T$ is Temperature in Kelvin, and $v$ is the velocity in cm/s. These choices roughly constrain each field to the range $(-1,1)$ such that its features are visible in linear space. The Ly-$\alpha$ flux is naturally in the range $(0,1)$ and does not require additional dedicated normalization. To constrain the output fields to their target ranges, we use a hyperbolic tangent filter on the final layer (ie. $x^\prime = \textrm{tanh}(x)$ ) for hydrodynamic fields and a sigmoid filter for Ly-$\alpha$. 

\begin{figure*}[t]
  \centering
  \includegraphics[width=\textwidth]{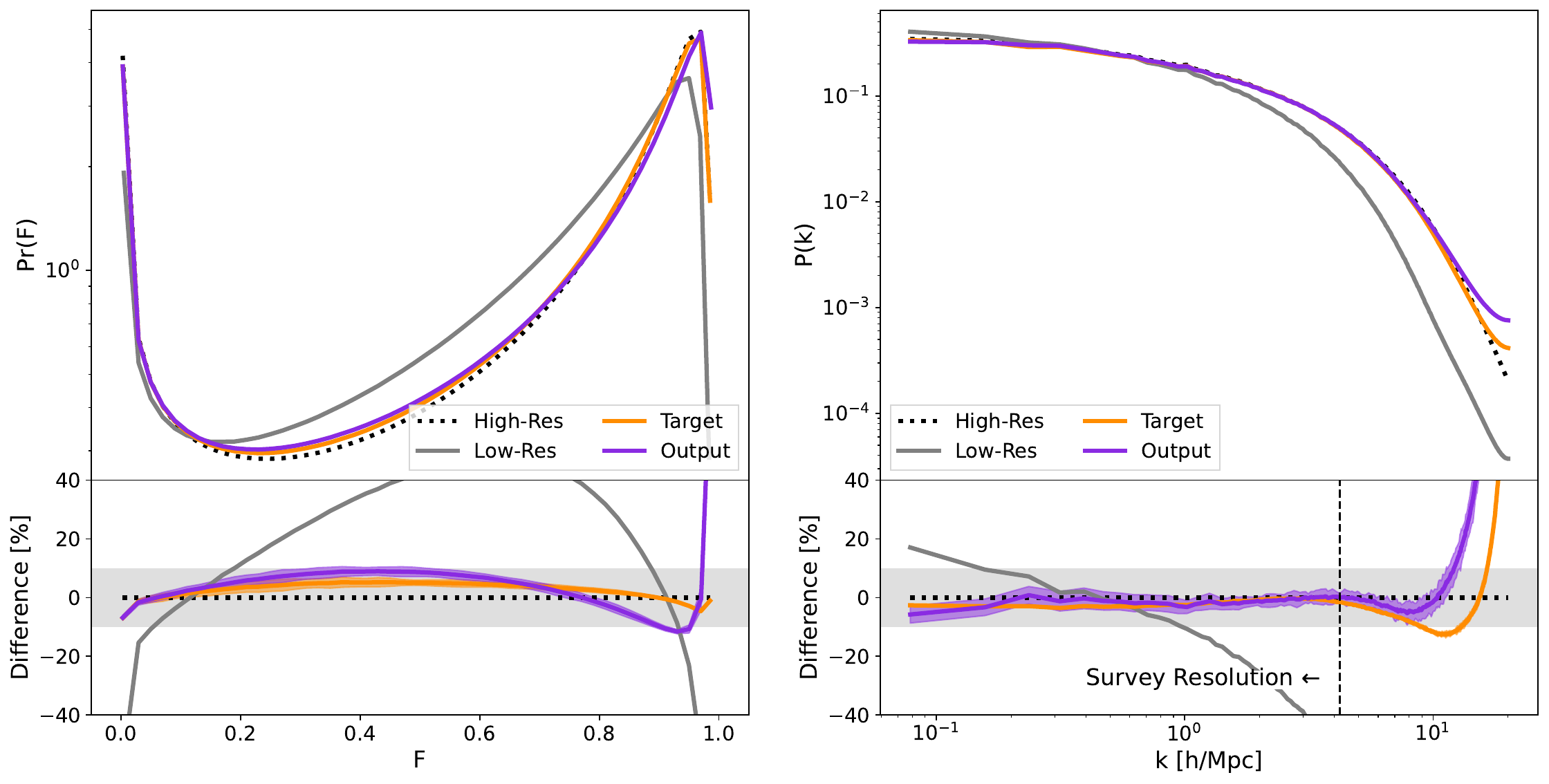}
  \caption{The Probability Density (left panel) and Power Spectrum (right panel) of Ly-$\alpha$ flux in Redshift space in four cases: the high and low-resolution Nyx simulations, the resized target, and our network output. The uncertainty band around our output shows the min-max range of an ensemble of 16 predictions. The grey band in the residual highlights the 10\% error range around the 'true' high-resolution target. Our output (in purple) shows clear improvement over the low-resolution Nyx simulation. }
  \label{fig:flux_stats}
\end{figure*}

\subsection{Loss Objective Design} \label{subsec:train_flux}

We employ multiple loss functions to incentivize our networks to reproduce various visual and statistical qualities present in the data. The dominant component of the total loss used for training is $\mathcal{L}_1$ distance, which encourages large-scale features to match between the predictions and targets, and improves training stability:
\begin{equation}
    \mathcal{L}_1 = \sum_{i=1}^{n} |y_i - f(x_i)|
\end{equation}
\noindent
Here, $y_i$ is the target (ie. `true') field which corresponds to the predicted field $f(x_i)$.

A well-known downside of $\mathcal{L}_1$ loss is its tendency to cause blurring of fine-scale features, so it is often augmented with a local or patch-based adversarial loss \citep{isola2018imagetoimage}. We adopt the well-tested approach from image synthesis applications of using a multiscale adversarial loss $\mathcal{L}_{\textrm{adv}}$ \citep{wang2018pix2pixHD,park2019semantic,TSIT}, which is defined by multiple patch-based discriminator networks that each see different resized versions of the generated and ground truth data. We refer the reader to \cite{wang2018pix2pixHD} for a full description and analysis of this conFiguration; we use their Eq. (5), including the feature matching terms, as our full adversarial loss $\mathcal{L}_{\textrm{adv}}$. More details on the discriminators are given in Appendix \ref{sec:hyperparam}.

Additionally, since the primary goal of the generator is to reconstruct statistics as accurately as possible, we introduce a spectral loss term, $\mathcal{L}_{\textrm{fft}}$ in addition to traditional $\mathcal{L}_1$ loss. $\mathcal{L}_{\textrm{fft}}$ is a measure of the difference between the Fourier transforms of the prediction and target (a stand-in for the power spectrum):

%Loss

\begin{equation}
\begin{gathered}
  \mathcal{L}_{\textrm{fft}} = (\sum_{i=1}^{n} |\widehat{y_i} - \widehat{f}(x_i)|^2)^{\frac{1}{2}} \\
\end{gathered}
\end{equation}

Here, $(\widehat{f})$ denotes the 3D Fast Fourier Transform, truncated to frequencies below 10 h/Mpc.
The complete loss for the generator is defined:

\begin{equation}
\mathcal{L}_{G} = \mathcal{L}_{\textrm{adv}} + \lambda_{\textrm{1}} \mathcal{L}_{1}  + \lambda_{\textrm{fft}} \mathcal{L}_{\textrm{fft}}
\end{equation}

\noindent
where $\lambda_1 = 100$ and $\lambda_{\textrm{fft}} = 6$ are hyperparameters to weigh the influence of $\mathcal{L}_{1}$ and $\mathcal{L}_{\textrm{fft}}$. After weighting, the relative scales of each of our loss terms are roughly: 

\begin{equation*}
\mathcal{L}_{1} \sim 20, \; \;
\mathcal{L}_{\textrm{fft}} \sim 10, \; \;
\mathcal{L}_{\textrm{GAN}} \sim 5, \; \;
\mathcal{L}_{\textrm{feat}} \sim 10.
\end{equation*}
 
\noindent
When $\lambda_1$ was set much lower, we observed that the adversarial model would occasionally infer realistic-looking volumes, but ones that did not match the ground truth as closely.

\section{Results} \label{sec:results}

To evaluate our model, we apply it to a separate volume at the same redshift that was not used for training (and with different random realization of the initial conditions than those used for the training volume). As is visually apparent in Figure \ref{fig:all_comp}, the reconstructed fields inferred by our model for this test simulation contain much sharper features than the low-resolution simulation, and the dynamic range of each field resembles the high-resolution field much more closely. At very small scales, we can observe both our model outputs and the resized target fields used to train them contain minor differences with respect to the reference high-resolution simulation, which we take to be the ``ground-truth''.

In the following sections, we will analyze in detail the statistical properties of our reconstructed fields and assess how closely their summary statistics match those of the test simulation, using a variety of probes. We focus first on our model's direct predictions of Ly$\alpha$, evaluating statistical properties of individual samples as well as ensemble predictions for quantifying model uncertainty. Finally we also evaluate some statistical properties of the underlying hydrodynamic fields predicted by our model.

\subsection{Lyman-alpha Fields} \label{subsec:lya_fields}

We show a visualization of the reconstructed Lyman-$\alpha$ field in Figure \ref{fig:all_comp}. This example 2D slice runs parallel to the line of sight, so that the redshift space distortions are visible along the x-axis. 

As seen in Figure \ref{fig:all_comp}, the reconstructed Lyman-$\alpha$ field is visually of high spatial fidelity and closely agrees with the resized target field. Agreement is especially good in diffuse regions and filaments, where densities are lower and there is more uniform motion along the line of sight. Conversely, the dense, hot regions surrounding galaxy clusters seem to be more difficult to handle, since the redshift space distortions amplify any discrepancies between the low-resolution and high-resolution simulations. These dense galaxy clusters make up an exceedingly small fraction of space by volume, so our model may struggle due to a relative lack of training data. 

The most commonly used statistics in evaluating accuracy of Lyman-$\alpha$ flux are the power spectrum (P(k)) and probability density function (PDF). We use the \texttt{ gimlet} code~\citep{Friesen2016} to calculate these statistics, as well as the P(k) and PDF of the baryon hydrodynamic fields (density, velocity, temperature). These statistics are calculated assuming sightlines or ``skewers" which pass through the entire volume, parallel to one of the axes.

\subsubsection{The Mean and Probability Density Function} \label{subsec:PDF}

The mean flux reconstructed by our model is only 2\% lower than the high-resolution target. In the following analysis we rescale all simulations to the same mean flux $\langle F \rangle = 0.6835$. This is the mean flux calculated from our high resolution training simulation, which is also in agreement with the observational measurements \citep[e.g.][]{Becker2013, Gaikwad2021}.

The probability density function, $Pr(F)$ is defined such that
the integral of $P$ over the full range of $F$ is equal to 1, 
$\int_{0}^{1} (F) dF = 1$. We show the PDF of our model's Ly$\alpha$ predictions compared against that of the high-res and low-res simulations in Figure \ref{fig:flux_stats}. We find that the PDF of our reconstructed Ly$\alpha$ is within 10\% of the true value and thus offers a substantial improvement over the low-resolution simulation, whose PDF is off by as much as 50\%.

\begin{figure}[t]
  \centering
  \includegraphics[width=\columnwidth]{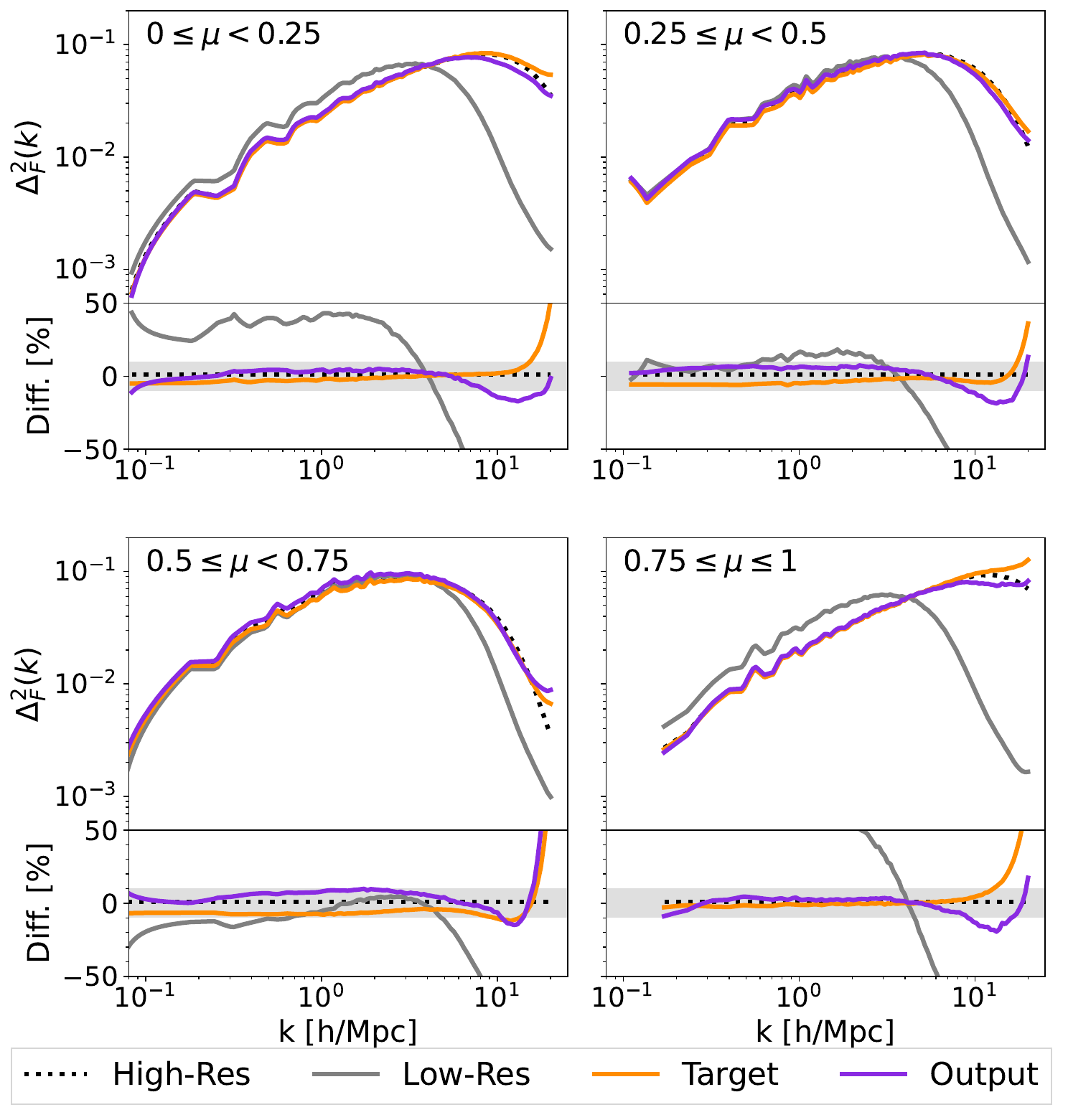}
  \caption{The dimensionless 3D flux power spectrum in 4 $\mu$ bins. The top left is mostly parallel to the line of sight, and the bottom right panel shows the power spectrum mostly perpendicular to the line of sight. The grey residual band highlights the 10\% error range.}
  \label{fig:pkmu}
\end{figure}

\subsubsection{The 1D Power Spectrum} \label{subsec:1dpk}

The power spectrum of the Ly-$\alpha$ forest offers a promising route to measuring both properties of the IGM and cosmological parameters. The 1D power Spectrum (P1D) is one of the most important observables that can be predicted by our cosmological models. We define the flux over-density as: 

% Flux overdensity
\begin{equation}
\delta_F(x) = (F(x)-\langle F \rangle)/\langle F \rangle
\end{equation}

\noindent
where $\langle F \rangle$ is the mean flux with respect to the entire volume in redshift space. In case of P1D, the power spectrum along a given skewer is given by:

% Fourier
\begin{equation}
\tilde{\delta}_F(k) = \int \delta_F(x) e^{ikx} dx
\end{equation}

% Fourier
\begin{equation}
P(k) =  \tilde{\delta}_F(k) \tilde{\delta}^{\ast}_F(k) 
\end{equation}

\noindent
where $\tilde{\delta}^{\ast}_F$ is the complex conjugate of $\tilde{\delta}_F$.
We plot the mean power spectrum $\langle P(k) \rangle$ averaged over modes with magnitude k from all skewers, as done in  \cite{Lukic2014}. 

\begin{figure*}[t]
  \centering
  \includegraphics[width=0.99\textwidth]{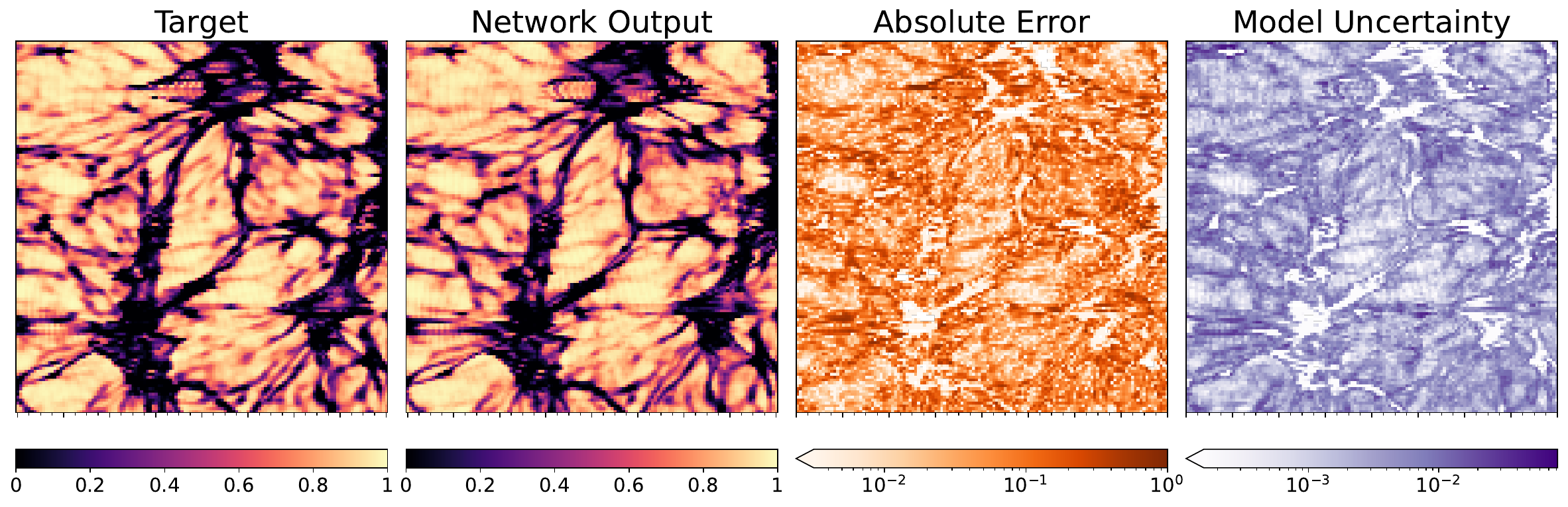}
  \caption{Comparison between uncertainty (fourth panel) and ensemble prediction error (third panel) Model Uncertainty is defined as the width of the interquartile range of an ensemble of 16 predictions. The error is defined as the absolute difference between the target field (first panel) and the ensemble mean (second panel). The regions of large uncertainty of an ensemble of outputs coincide with the regions of greatest prediction error.}
  \label{fig:ensemble}
\end{figure*}

We show comparisons between the generated and true P1D in Figure \ref{fig:flux_stats}. Our reconstructed Ly$\alpha$ P1D is in agreement with the high resolution test volume to within 5\% for wavenumbers smaller than 10 h/Mpc, after which the resized target we trained the model on begins to diverge due to aliasing. Although accurate reconstruction of P1D is not the main target of our method -- given that it can be reproduced by the ``brute force'' hydrodynamical simulations -- it is very reassuring that this important statistics which used in so many inference works \citep[][, to mention a few]{Boera2019, Walther2019, PDB2020, Rogers2020, Walther2021} can be precisely reconstructed from otherwise quite inaccurate low-resolution hydrodynamical simulations which yield the incorrect P1D even at low wavenumbers for Ly$\alpha$.

\subsubsection{The 3D Power Spectrum} \label{subsec:3dpk}

Unlike the hydrodynamic fields, the flux field is distorted along the line of sight due to redshift-space distortions. It is therefore not appropriate to average out flux power in all 3 dimensions due to this anisotropy. Instead of averaging over spherical shell bins of wavenumber $k$, a common method of describing an anisotropic power spectrum is in terms of $k$ and $\mu$, the angle between the wavevector $\bf k$, and its LOS component:

\begin{equation} \label{eq:pkmu}
\mu \equiv k_{\parallel}/|\bf{k}|
\end{equation}

Therefore, the 3D power spectrum (P3D) we present is binned in k and $\mu$, rather than $k$ alone, resulting in $P(k,\mu)$. We separate this family of power spectra into 4 bins in $\mu: [0,0.25), [0.25, 0.5), [0.5, 0.75), [0.75, 1]$. The later bin is most nearly parallel to the line of sight, while the first bin is almost perpendicular to the line of sight.

We show the 3D power spectrum in these 4 bins in Figure \ref{fig:pkmu}. For all 4  $\mu$ bins, our reconstructed Ly$\alpha$ achieves is within $<10\%$ of the true P3D for wavenumbers smaller than 10 h/Mpc. The improvement is greatest in the 4th bin, $0.75 \le \mu \le 1$, where the low resolution is off by $>50\%$ for most scales. This bin is also the most important for line-intensity mapping since it represents the spectral power most nearly along the line of sight. The substantial accuracy improvements offered by our reconstructed P3D opens the door for building accurate Ly-$\alpha$ mock skies in Gpc volumes, using 4096$^3$-8192$^3$ grids.

\begin{figure}[b]
  \centering
  \includegraphics[width=1\columnwidth]{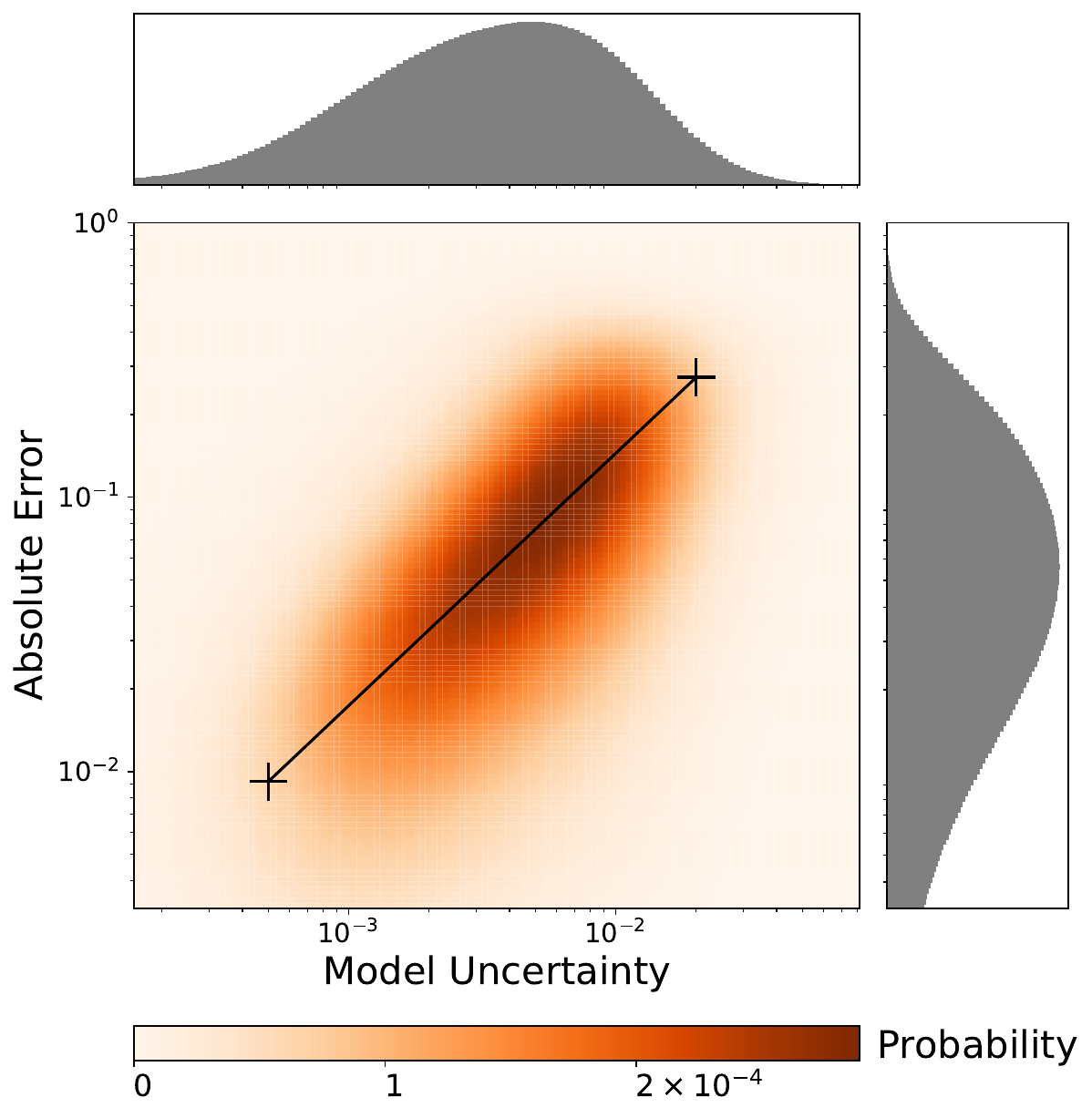}
  \caption{Correlation between the model uncertainty (interquartile range) of an ensemble of reconstructions and error of their mean in redshifted flux. The line shows a best-fit power-law with an exponent of 0.92.}
  \label{fig:Variance}
\end{figure}

\subsection{Uncertainty Quantification} \label{sec:hydro noise}

As described in Section \ref{sec:hydro noise}, we generate probabilistic ensemble forecasts by incorporating noise into the network at multiple scales. This approach allows for sampling repeated predictions to form a probabilistic ensemble forecast for a single region of space, even when using a single set of input fields, which gives us a measure of the uncertainty of our model's prediction.

Figure \ref{fig:ensemble} presents an example ensemble forecast, illustrating the mean and interquartile range per grid-cell across a set of 16 unique ensemble members. Qualitatively, the results demonstrate higher variability (uncertainty) in challenging locations such as the edges of absorption features, and lower variability in the center of large transmission or absorption regions. This aligns with expectations, as these areas adjacent to regions of zero flux are the most sensitive to small changes in the hydrodynamic fields and the discrepancy between low- and high-resolution simulations. We chose to use the interquartile range to represent the model uncertainty as the distribution is non-Gaussian, but using the variance or min-max range produced similar correlations.

As shown in Figure \ref{fig:ensemble}, the predictive error of the ensemble mean is greatest in roughly the same regions where the variability is greatest. Figure \ref{fig:Variance} shows the cell-wise correlation between the true error (which we know for the case of test data) and the variability of the ensemble as measured by the interquartile range. The error seems to correlate with the variance according to a power-law, meaning we can safely treat the variance as a surrogate for the cell-wise error. 

This suggests that it is possible to use this ensemble variability to estimate the uncertainty associated with cell-wise flux prediction.  This uncertainty can be later be propagated through the scientific pipeline as needed.

\subsection{Hydrodynamic Fields} \label{subsec:Hydro_fields}

Inspecting the Baryon Density fields in Figure \ref{fig:all_comp} by eye, we see that our reconstruction matches the target field almost everywhere. The most prominent features missing from the Baryon Density reconstruction are the smallest filaments, especially those nearest the large clusters. These filaments are so thin that they are only a couple grid-cells wide in the target field and are often not visible at all in the low-resolution simulation, making them challenging to model.

\begin{figure}[t]
  \centering
    \includegraphics[width=\columnwidth]{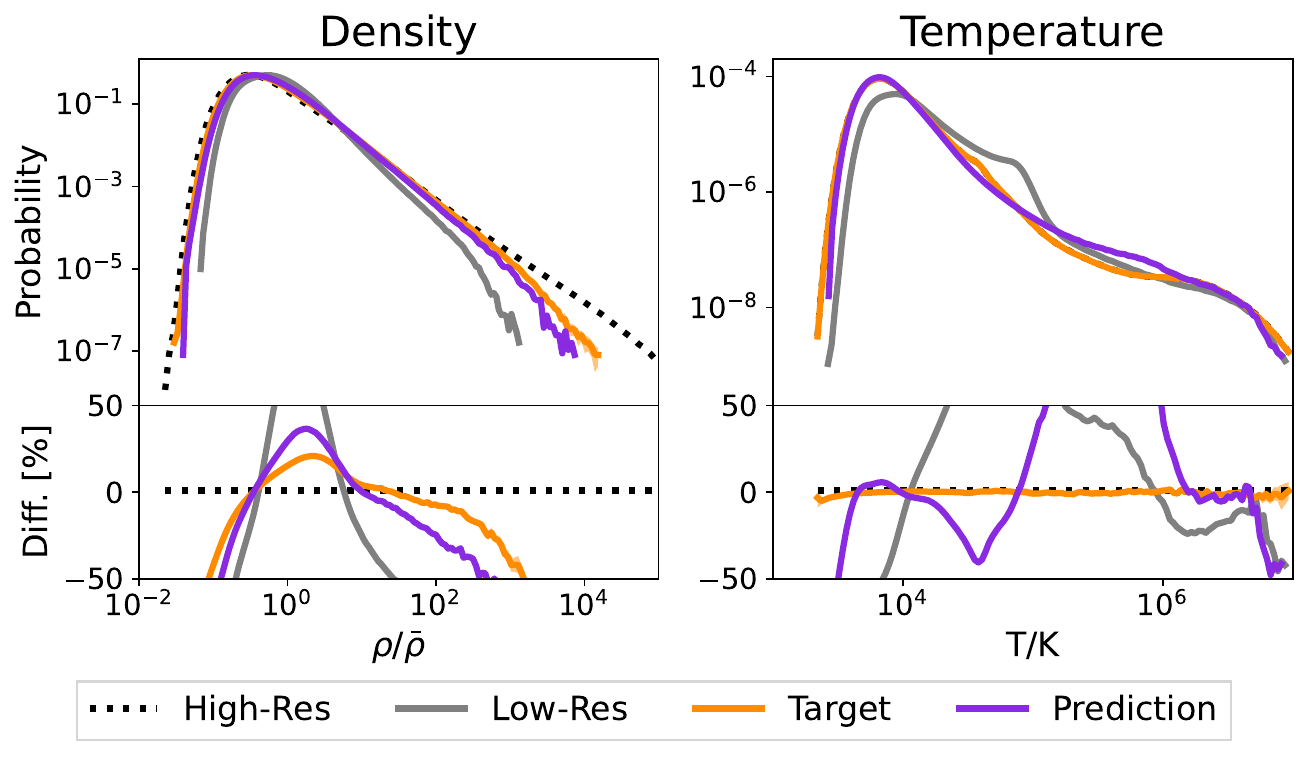}
  \caption{The Probability Density Functions of the predicted density and temperature fields compared against the reference distributions from the training target and high-resolution simulation. Reconstructing hydrodynamic fields is not the focus of our method, but nevertheless we provide improvement over the low-resolution fields.}
  \label{fig:Hydro_stats}
\end{figure}

\begin{figure}[b]
  \centering
    \includegraphics[width=\columnwidth]{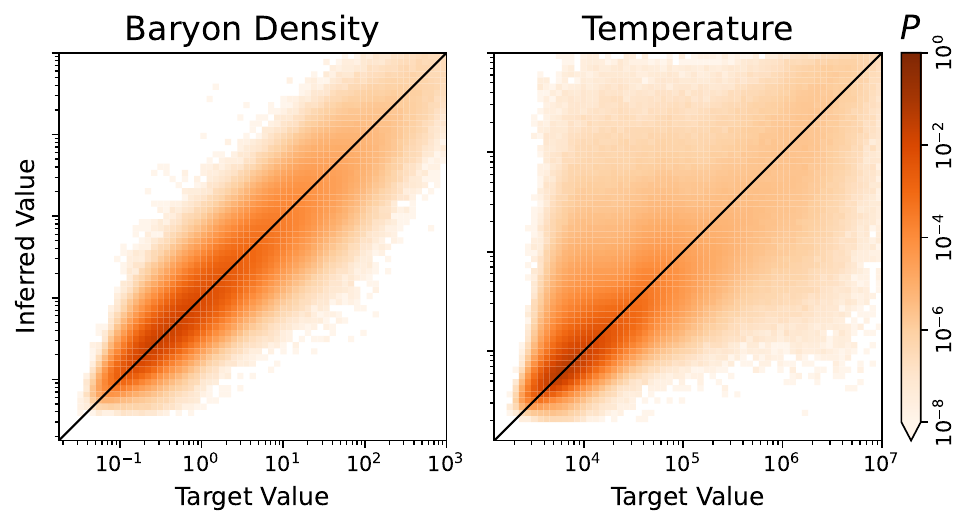}
  \caption{Cell-wise correlation between the reconstructed and target density and temperature fields. The diagonal line shows one-to-one correspondence. Our model struggles with the high temperature shock regions so the spread is wider in the high end of temperature range.}
  \label{fig:correlation}
\end{figure}

\begin{figure*}[t]
  \centering
  \includegraphics[width=\textwidth]{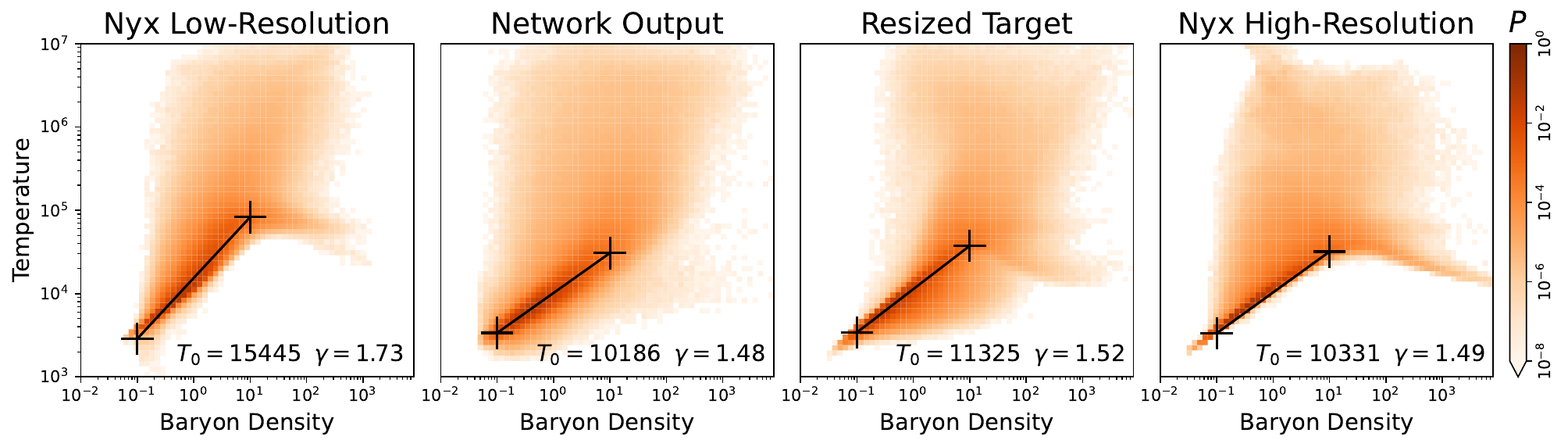}
  \caption{The density–temperature distribution of the gas in our model predictions compared against the reference distributions from the low-resolution, high-resolution, and resized simulations. For each output we plot the best-fit power-law relationship between density and temperature, and we find our model provides the most accurate $T_0$ and $\gamma$ estimates when compared against the high-resolution simulation.} 
  
  \label{fig:2D_hist}
\end{figure*}

For most regimes, including the regime relevant for the IGM and Ly-$\alpha$ forest, the reconstructed temperature field is accurate as well. However, our model also struggles to capture the complexity of thermal shocks surrounding galaxy clusters, such as the two bright points in Figure \ref{fig:all_comp}. These turbulent features result from the accretion of matter from filaments onto galaxy halos and are commonly found around large gravitationally collapsed structures. The high-resolution simulation is able to more accurately capture the complex fluid dynamics of these shocks, but the low-resolution simulation cannot capture this detail and blurs them into much wider and more diffuse clouds. Translating between these diffuse clouds to complex, turbulent shocks presents a challenge due to the difference in information content of the two fields. Our model is able to shrink the diameter of the reconstructed shock closer to the target, but is unable to fully represent the turbulent details, which are not present in the input data.

As shown in Figure \ref{fig:Hydro_stats}, the PDF of baryon density reconstructed by our model is closer to the target field than the input field virtually everywhere. The PDF also reveals that our reconstructed Baryon Density field is notably much more diverse than the low-resolution input, as it contains grid cells with more extreme values than it's input (both more dense and more diffuse), making it even more similar to the target field.

The agreement of the PDF for temperature is less impressive than for density. Even so, our model offers substantial improvement over the low-resolution Nyx simulation, particularly in the tails of the temperature PDF.

\subsubsection{Gas Phase Statistics} \label{subsec:IGM}

The properties of the gas which makes up the intergalactic medium (IGM) have been well studied, and we wish to compare our reconstructed results to the expected statistics. We hope to be able to faithfully reproduce the phase distribution of the IGM. 

We plot the density-temperature distribution in Figure \ref{fig:2D_hist}.  
As a result of the downsampling scheme used to generate our targets (see Appendix \ref{sec:downsampling}), much of the diversity in the gas density–temperature distribution is reduced in our reconstruction inference. The physics of our simulations do not capture processes in the condensed galaxy-forming regions and may not have enough statistical accuracy in the warm-hot intergalactic medium. However, those regimes are of little consequence for the Ly-$\alpha$ forest. Our method does preserve the power-law relationship of the diffuse gas:
% IGM power law
\begin{equation} \label{eq:IGM}
T = T_0 (\rho / \langle \rho \rangle)^{\gamma-1}
\end{equation}
Here $T_0$ is a lower end of the power-law and $\gamma$ is the slope.
The regime the follows this power-law represents the majority of the gas and is by far the most important for predicting the Ly-$\alpha$ flux.  

This power law is fit using \texttt{gimlet} code and the resulting parameters for each volume are shown in Figure \ref{fig:2D_hist}. The power-law parameters produced by the low-resolution simulation are off by more than 10\%, while our method reconstructs these parameters to within about 1\%. Measurements of the $T_0$ and $\gamma$ for different redshifts provide insights into the thermal and ionization history of the IGM, as both H and HeII reionizations are expected to significantly affect these parameters \citep[see, for example,][]{Walther2019, Gaikwad2021}.

The region of phase space below this power law is unpopulated in both low and high resolution simulations but is populated in our reconstructed result and in the target volume used for training. This erroneous population is introduced through some artifact of our re-sampling scheme (as described in Appendix \ref{sec:downsampling}). We have chosen a methodology that best preserves the spectral qualities of the high-resolution data but introduces some more subtle statistical artifacts like this. In future work we hope to get around this issue, but since we are using a target which is shrunken by a factor of 8, it is not realistic to expect that all statistics can be preserved. This erroneous population accounts for very little of the total volume of our reconstructed model and has little impact on the properties of the derived lyman-$\alpha$ fields.

\section{Conclusion} \label{sec:conclusion}

In this work, we have presented a generative deep learning method for translating between low and high-resolution simulations of hydrodynamic structure formation. We have demonstrated that this translation provides an accurate reconstruction over a wide range of scales and captures a number of important statistical properties. Our model output can match the 'true' statistical properties of the target hydrodynamical fields in the regimes relevant to the observables, and outperforms the low-resolution input simulation in all regimes. The observable, Ly-$\alpha$ flux, derived from our model output is within a few percent of the 'real' simulation's powerspectra up until scales of around $\sim$ 10 h/Mpc. 

\begin{figure*}[t]
  \centering
  \includegraphics[width=\textwidth]{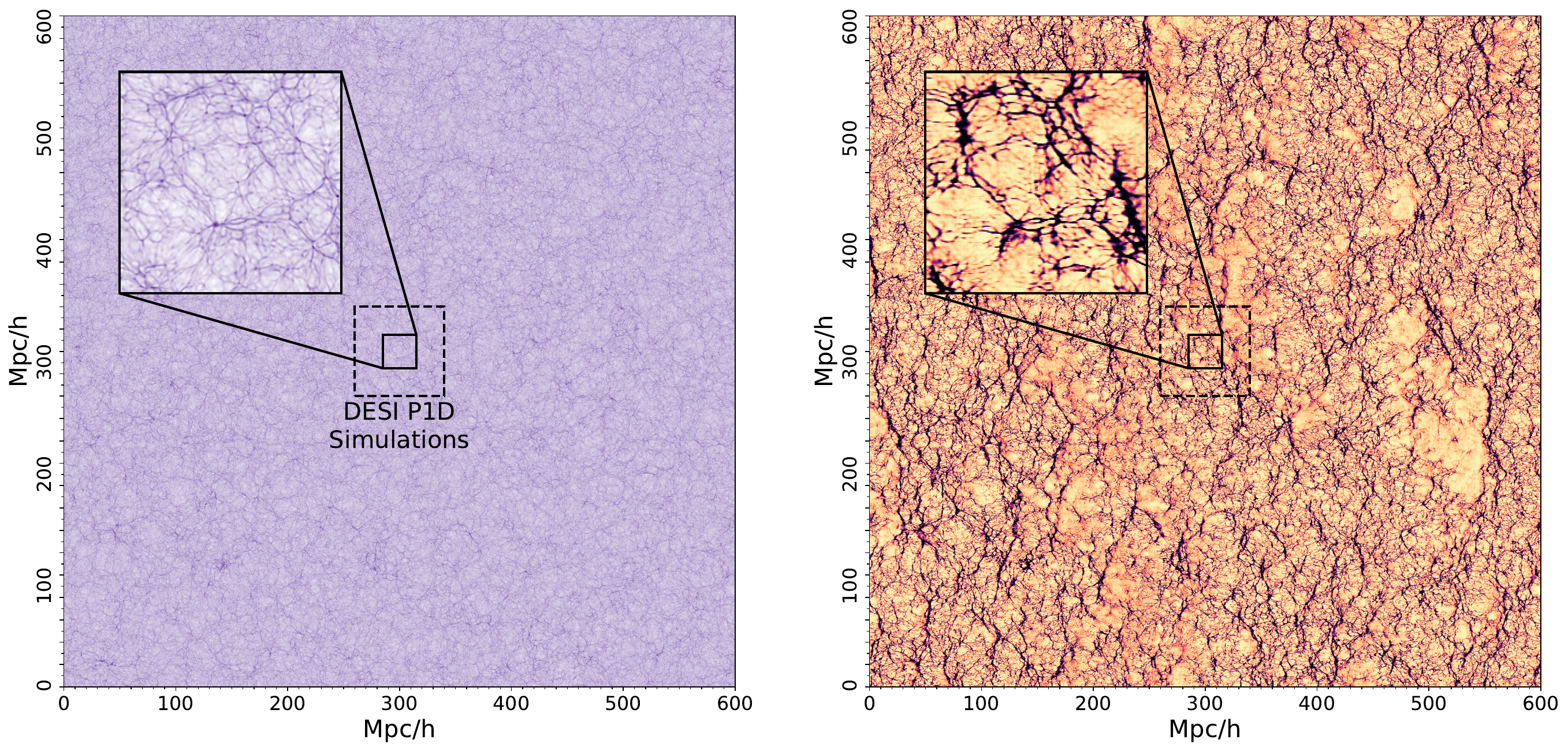}
  \caption{Visualization of the Baryon Density and Flux reconstructed by our model on a 600 Mpc/h wide volume. The dashed box highlights a region 80 Mpc/h wide, which is the size of our training data.}
  \label{fig:big_boy}
\end{figure*}

The scales accurately captured by this method go well beyond the resolution of the DESI survey \citep{Walther_2021}. Our method provides a far less computationally expensive alternative to high-resolution hydrodynamic simulations when generating mock survey data. 

Our network can be conveniently trained on a relatively small volume for some given cosmological and astrophysical parameters, and then used to reconstruct an arbitrarily large volume. As shown in Figure \ref{fig:big_boy}, even an 80 Mph/h sample is sufficient to accurately train the model. This feature enables significant computational savings when generating larger mock catalogs. 

Using noise injection in our generator, we can bridge the difference in the spectral power and information content of the input and target volumes. However, we also demonstrate that this noise injection allows us to better understand the predictive uncertainty of the reconstructed outputs. The cell-wise diversity of an ensemble of reconstructed outputs correlates with regions of greater inference error. This provides us with a means of estimating the uncertainty of our output which can be propagated through the cosmological analysis pipeline.

There are some notable shortcomings in the hydrodynamic field reconstructions, particularly in and around the larger clusters and shocks. While these regions represent a very small percentage of the universe by volume and have only marginal effect on the Ly-$\alpha$ statistics used for cosmology, they are of significant astrophysical interest. Because these regimes are especially dependant on subgrid dynamics, they are poorly captured by the low-resolution simulation and present the greatest challenge to our machine learning model. However, our method of refining the hydrodynamic fields from low-resolution simulations presents a notable improvement over previous machine learning methods which reconstructed the hydrodynamic fields from n-body dark matter alone and could not capture these shock fronts at all \citep{Harrington_2022, Horowitz_2022}. Because we begin with realizations of the hydrodynamic fields, our model does not need to guess where these shocks are.

\section{Acknowledgments} \label{sec:Acknowledgments}

This work was partially supported by the DOE’s Office of Advanced Scientific Computing Research and Office of High Energy Physics through the Scientific Discovery through Advanced Computing (SciDAC) program. This research used resources of the National Energy Research Scientific Computing Center, a DOE Office of Science User Facility supported by the Office of Science of the U.S. Department of Energy under Contract No. DEC02-05CH11231. This research also used resources of the Argonne Leadership Computing Facility, which is a DOE Office of Science User Facility supported under Contract DE-AC02-06CH11357.

\newpage
%\section{References} \label{sec:conclusion}
%\onecolumn
\bibliography{main}
\bibliographystyle{aasjournal}



\appendix
\twocolumngrid
\section{Hyperparameters \& Architecture Details} \label{sec:hyperparam}

Our multiscale patch-based discriminator is based on the implementation of TSIT \citep{Jiang20}, which itself adopts this approach from \citep{wang2018pix2pixHD}. The adversarial loss term $\mathcal{L}_{\text{adv}}$ is given by a weighted combination of the standard min-max GAN loss \citep{Goodfellow2014} and feature matching loss \citep{salimans2016improved} to improve training stability for the generator (see Eq. (5) in \cite{wang2018pix2pixHD}). 

The multiscale discriminator is composed of four different discriminators that work on the same input features at varying spatial scales. One of the discriminators works on the full resolution input, while the others operate on input that has been progressively downsampled by factors of 2. The input for each discriminator includes the real or inferred hydrodynamic field and the input field fed to the generator, which are concatenated along the channel dimension to enhance training stability. Each discriminator produces a 3D grid of predictions that classify spatial patches of the input as real or fake, enabling the model to focus on local texture and fine-scale details.

We list other relevant network hyperparameters in Table \ref{tab:hp}.

\begin{table}[h]
    \caption{Hyperparameters for our netowrks.}
    \label{tab:hp}
    \centering
    % \footnotesize
    \begin{tabular}{ll}
    \toprule
    Hyperparameter & Value \\
    \midrule
    Global batch size & 16 \\
    Training crop size & $128^3$ \\
    Optimizer & Adam($\beta_1=0.5$, $\beta_2=0.999$) \\
    Learning rate & 5e-4 \\
    $\lambda_{L1}$ & 800 \\
    $\lambda_{\text{fft}}$ & 400 \\
    $\lambda_{\text{Feat}}$ & 1 \\
    Generator upsampling blocks & 6 \\
    Multi-scale discriminators & 4 \\
    Discriminator layers & 6 \\
    
    \bottomrule
    \end{tabular}
\end{table}

\vspace{2cm}

\section{Inference} \label{sec:inference}

Our model is fully convolutional and thus agnostic to input shape. This allows us to train it on smaller, more manageable volumes and perform inference on much larger ones. Training on smaller volumes is unavoidable, as we wish to perform inference on low-resolution simulations of hundreds of Mpc/h to a few Gpc/h on a side, but we can only accurately simulate high-resolution Ly-$\alpha$ flux in boxes of $\sim$80 Mpc/h. 

However, we cannot perform an inference on an entire volume at once. Large volumes may have as many as $6144^3$ cells, while a GPU running our neural network can handle at most around $256^3$ cells at a time. To infer a large volume, we break it up into much smaller ``chunks'' which can be inferred individually and in parallel.

The neural network has less information and context available to it when inferring the cells closest to the edge of an input volume, causing it to falter and predict erroneous features. This leads to ``edge effects'' which are noticeable artifacts and overall greater inference error around the edges of a predicted volume.

These edge effect artifacts would appear throughout the whole simulated volume, once the inferred chunks are stitched back together. To prevent this, we can make each chunk slightly wider than is needed and later trim off the overlapping regions. In practice, we tile the field with chunks of $256^3$ but add an extra 32 ``ghost'' cells to each inside face which are removed after inference. This greatly decreases the overall error and minimizes discontinuities between chunks.

\begin{figure}[b]
  \centering
  \includegraphics[width=0.8\columnwidth]{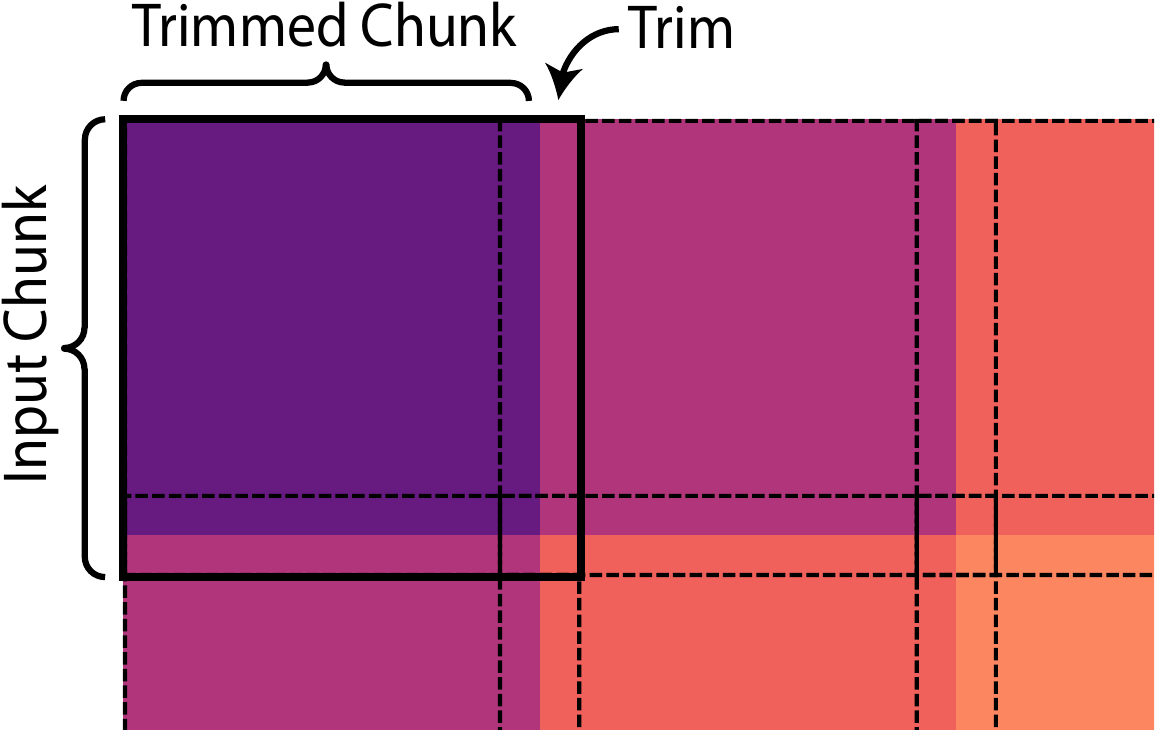}
  \caption{To satisfy memory limitations, inference is performed on small chunks of the desired volume, one at a time. To reduce edge effects, the chunks overlap with ghost cells which are removed after inference is performed.}
  \label{fig:Inference_chunks}
\end{figure}

\section{Resizing Methodology} \label{sec:downsampling}

Finding a method to resize large volumes presented a unique challenge. In usual image processing circumstances, different resizing methodologies do not result in vastly different outputs. However, our circumstance is unlike common image problems as our signal is populated by very thin features and spans many orders of magnitude; not to mention our application requires resizing by a factor of 8, so we must be especially cautious.  It is thus important that we preserve the statistics of our target fields as precisely as possible, and we experimented with several different resizing methods.

\subsection{Subsampling}
The most naive method for resizing a volume is to simply extract an evenly spaced subset of the high-resolution field. In our case, this means selecting one out of every 512 cells to represent that region of space. We sampled from about the center of that box using: 

\noindent
\texttt{field = field[3::8, 3::8, 3::8]}

This virtually guarantees that the probability distribution of each field's values will be preserved but can entirely disrupt the power spectrum. For subsampling, the overall power contained in the power spectra is preserved, so any power contained in scales above the new Nyquist frequency is aliased and disruptively shifted into lower wavenumbers \citep{approximation}. This distorts the power spectrum, especially in the case of Baryon Density where there would be much structure and spectral power on scales past the Nyquist limit.

\begin{figure}[t]
  \centering
  \includegraphics[width=\columnwidth]{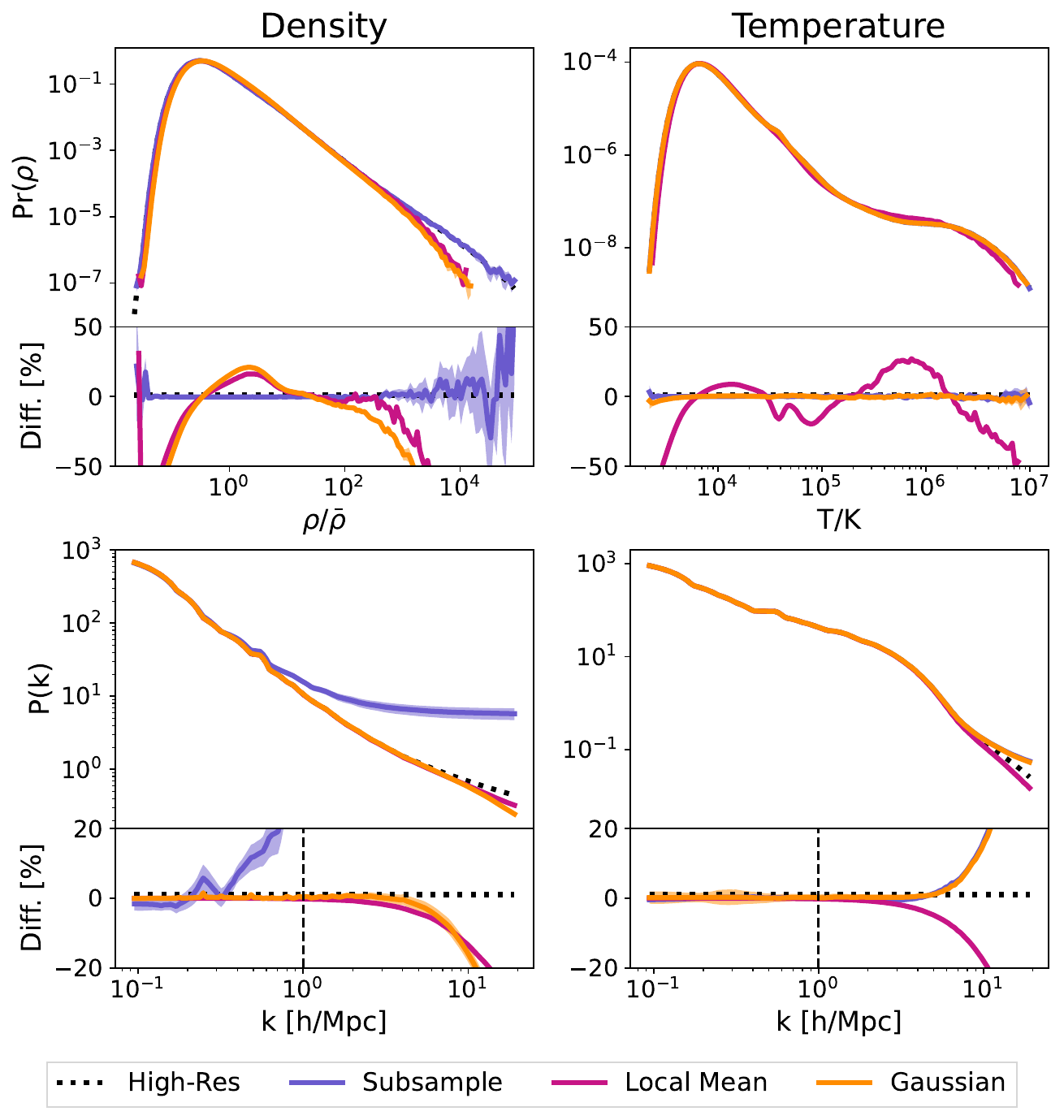}
  \caption{Hydrodynamic field statistics for each resize method applied to our validation field.}
  \label{fig:Resize_Hydro_stats}
\end{figure}

\subsection{Local Mean}
Another plausible candidate for resizing our fields is to take the mean of each local cube of $8^3$ cells using:

\noindent
\texttt{ skimage.transform.downscale\_local\_mean()} 

Unlike subsampling, this method preserves the power spectrum at small scales. It is not perfect, but because it smooths the fields as it shrinks, the local mean operation does not introduce erroneous jagged features. But because each local neighborhood is pushed towards its average, this method inevitably skews the probability distribution towards the most common values.  However, we prefer to minimize changes to both of these statistics.

\subsection{The Gaussian Pyramid} \label{subsec:gauss}
We can find a middle ground between the two methods above by combining a subsampling routine with a Gaussian blur. If the spatial reduction we required in each dimension was only a factor of 2 or so, a single low-pass and resample would likely suffice. However, for our purposes, we wish to reduce the volume's scale by a factor of 8 or more in each dimension. To achieve this while best preserving the field's power spectrum, we will apply multiple passes of a filter--then-resample scheme to gently resize our training data in a way which is more sensitive to the diverse scale space.

For a factor of 8, the Gaussian pyramid is executed by the following sequence: 

\begin{verbatim}
for i in range(3):
    field = scipy.ndimage.gaussian_filter(
            field, sigma=0.5, mode="reflect")
    field = field[::2, ::2, ::2]
\end{verbatim}

\subsection{Comparison} \label{subsec:comp}

We plot the probability density functions and power spectra for each of these methods in Figure \ref{fig:Resize_Hydro_stats}. The subsampling and pyramid methods can be applied in 8 unique ways (choosing from 8 central cells for subsampling), so we can establish a kind of dispersion shown as the shaded regions in the plot. We choose to use the pyramid method to train on as the subsampling routine distorts the baryons density's power spectra, and the local mean distorts the temperature PDF. 

The Orange line, representing our Gaussian pyramid method, is close to the high resolution statistics in all cases. Sub-sampling (blue), may better preserve the PDF, but introduces wild errors of over an order of magnitude in the power spectrum. Because of this, we finally settle on the Gaussian scheme to generate our training data.

\end{document}